\documentclass[12pt]{iopart}

\usepackage{graphicx}
\usepackage{epsfig}             % to insert PostScript figures

\begin{document}

%%%%%% Title of paper %%%%%%%%

\title{Optimum Small Optical Beam Displacement Measurement}

%\author{Magnus T. L. Hsu}

%\author{Vincent Delaubert} 

%\author{Ping Koy Lam} 
%\email[Email: ]{ping.lam@anu.edu.au}

%\author{Warwick P. Bowen} 

%\affiliation{Australian Centre for Quantum-Atom Optics, Department 
%of Physics, Australian National University, ACT 0200, Australia}

\author{Magnus~T.~L.~Hsu, Vincent~Delaubert, Ping~Koy~Lam\footnote[3]{To whom correspondence should be
addressed (ping.lam@anu.edu.au)}~and Warwick~P.~Bowen}

\address{Australian Centre for Quantum-Atom Optics, Department
of Physics, Australian National University, Canberra ACT 0200,
Australia.}

\date{\today}

%%%%%% Abstract %%%%%%%%%%%
\begin{abstract}
We derive the quantum noise limit for the optical beam displacement of
a TEM$_{00}$ mode.  Using a multimodal analysis, we show that the
conventional split detection scheme for measuring beam displacement is
non-optimal with $\sim 80$~\% efficiency.  We propose a new
displacement measurement scheme that is optimal for small beam
displacement.  This scheme utilises a homodyne detection setup that
has a TEM$_{10}$ mode local oscillator.  We show that although the
quantum noise limit to displacement measurement can be surpassed using
squeezed light in appropriate spatial modes for both schemes, the
TEM$_{10}$ homodyning scheme out-performs split detection for all
values of squeezing.
\end{abstract}


% Uncomment for Submitted to journal title message
\submitto{\JOB}

%\maketitle must follow title, authors, abstract, \pacs, and \keywords
\maketitle

%%%%%%%% Introduction %%%%%%%%%%
\section{Introduction}
Efficient techniques for performing optical beam displacement
measurements are crucial for many applications.  When an optical beam
is reflected from, or transmitted through, an object that is moving,
the mechanical movement can be translated to a movement of the optical
beam.  Characterisation of the transverse position of this beam then yields an extremely accurate measurement of the object movement.  Some example applications that use these techniques are: {\it Atomic force
microscopy}, where a beam displacement measurement is used to
characterise the vibration of a cantilever, and the force the
cantilever experiences \cite{AFM, putman}; {\it inter-satellite
position stabilisation}, where a displacement measurement allows a
receiving satellite to orient itself to an optical beam sent by
another satellite, thus allowing a reduction of non-common mode
positional vibrations between satellites \cite{arnon, nikulin}; and
{\it optical tweezer}, where the position of particles held in an
optical tweezer can be detected and controlled by measuring the
position of the beam \cite{guo, simmons, gittes, denk}.  An
understanding of the fundamental limits imposed on these
opto-mechanical positional measurements is therefore important.

Recently there has been increasing interests, both theoretical
\cite{fabre} and experimental \cite{treps1d, treps2d, treps2da}, in
using quantum resources to enhance optical displacement measurements.
Much of the interest has been on how multi-mode squeezed light can be
used to enhance the outcome of split detector and array detector
measurements.  This is an important question since split detectors and
arrays are the primary instruments presently used in displacement
measurements and imaging systems.

In spite of the successes in using multi-mode squeezed light to
achieve displacement measurements beyond the quantum noise limit
(QNL), we will show in this paper that split detection is not an optimum displacement measurement.  Assuming that the beam under interrogation is a TEM$_{00}$ beam, we perform a multimodal analysis to derive the QNL for optical displacement measurements.  We then analyse split detection, the conventional technique used to characterise beam displacement, and compare it to the QNL. We find that displacement measurement using split detection is not quantum noise limited, and is at best only $\sim 80$~\% efficient.  As an alternative, we consider a new homodyne detection scheme, utilising a TEM$_{10}$ mode local oscillator beam.  We show that this scheme performs at the QNL in the limit of small displacement.  This technique, which we term TEM$_{10}$ homodyne detection, has the potential to enhance many applications presently using split detectors to measure displacement. Furthermore, the QNL for optical displacement measurement can be surpassed by introducing a squeezed TEM$_{10}$ mode into the measurement process.

%%%%%%% Displacement Measurement %%%%%%
\section{Displacement Measurement}

\subsection{Quantum Noise Limit}

The position of a light beam can be defined as the mean position of
all photons in the beam. Beam displacement is then quantified by the amount of deviation of this mean photon position from some fixed reference axis. In this paper, we assume that the displaced beam has a transverse TEM$_{00}$ mode-shape.  To simplify our analysis, we assume, without loss of generality, a one-dimensional transverse displacement $d$ from the reference axis.  The normalised transverse beam amplitude function for a displaced TEM$_{00}$ beam, assuming a waist size of $w_{0}$, is given by
\begin{equation} \label{disp00}
u_{0}(x-d) = \left(\frac{2}{\pi w_{0}^{2}} \right)^{1/4}\exp
\left[ - \left( \frac{x-d}{ w_{0}} \right)^{2} \right]
\end{equation}
The transverse intensity distribution for a beam with a total of $N$
photons is then given by $\langle I\rangle = Nu_{0}^{2} (x-d)$.  This
equation essentially describes the normalised Gaussian spatial
distribution of photons along one transverse axis of the optical beam.

For a coherent TEM$_{00}$ light beam, the photons have Gaussian distribution in transverse position, and Poissonian
distribution in time.  It is clear that a detector which discriminates
the transverse position of each photon will provide the maximum
possible information about the displacement of the beam.  Such
discrimination could, for example, be achieved using an infinite
single photon resolving array with infinitesimally small pixels.
Although in reality such a detection device is unfeasible, it
nevertheless sets a bound to the information obtainable for beam
displacement without resorting to quantum resources.  This bound
therefore constitutes a quantum noise limited displacement
measurement.  More practical detection schemes can therefore be
benchmarked against this limit.

Let us now examine an optimum measurement of beam displacement using
our idealised array detector.  Using equation (\ref{disp00}), the
probability distribution of photons along the $x$-axis of the detector
is given by
\begin{equation}
P(x)=\sqrt{\frac{2}{\pi w_{0}^{2}}} \exp \left[ -2 \left(
\frac{x-d}{ w_{0}} \right)^{2} \right]
\end{equation}
As each photon in the beam impinges on the array, a single pixel is
triggered, locating that photon.  The mean arrival position of each
photon $\langle x \rangle$ and the standard deviation $\Delta d$ are 
given by
\begin{eqnarray}
\langle x \rangle &=&\int^{\infty}_{-\infty}xP(x)dx = d, \\
\Delta d &=&\sqrt{\int^{\infty}_{-\infty}x^{2}P(x)dx -
d^{2}}=\frac{ w_{0}}{2}.
\end{eqnarray}
From the arrival of a single photon we can therefore estimate the
displacement of our mode with a standard deviation given by $\Delta
d$.  For $N$ photons, the standard deviation becomes $\Delta d_{\rm
QNL}=\Delta d/\sqrt{N}$.  The minimum displacement discernible by a given detection apparatus is directly related to the sensitivity of the apparatus, defined here as the derivative of the mean signal divided by its standard deviation.  For infinite array detection, the signal expectation value is equal to the displacement.  Thus the QNL for optimal displacement measurement sensitivity is given by
\begin{equation} \label{arrayslope}
   {\cal S_{\rm QNL}} = \frac{1}{\Delta d_{\rm QNL}} = \frac{2\sqrt{N}}{w_{0}}
\end{equation}
A plot of $\cal {S_{\rm QNL}}$ as a function of displacement is shown
in Fig.~\ref{SNR}.  Since all pixels in the array are assumed to be
identical, the standard deviation $\Delta d_{\rm QNL}$ is independent
of displacement.  We therefore observe that $\cal{S_{\rm QNL}}$ is
constant for all displacements.  $\cal{S_{\rm QNL}}$ is also
proportional to $\sqrt{N}$ due to the quantum noise limited nature of
the beam.  Finally, the inverse scaling with waist size $w_{0}$
suggests that the accuracy of a displacement measurement can be enhanced by focussing the beam to a smaller waist.

\subsection{Split Detection}

\begin{figure}[!ht]
\begin{center}
\includegraphics[width=6cm]{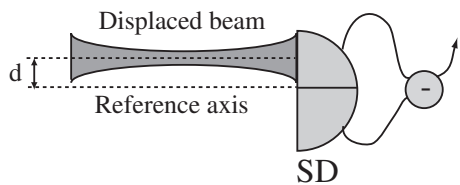}
\caption{Split detection.  The difference photo-current of
the two halves measures the beam displacement, $d$.}
\label{SD}
\end{center}
\end{figure}

In the previous section, we saw how an idealised array detector can be
used to perform quantum noise limited displacement measurements.
Implementation of such a detector, however, is clearly impractical.
The most common technique for displacement measurement is {\it
split detection} \cite{putman, treps1d, treps2d}.  In this
scheme, the beam under interrogation is incident centrally on a split
detector.  The difference between the two photo-currents of the two
halves then contains information about the displacement of the beam
(see Figure \ref{SD}).

At this stage we must introduce a more methodical representation of
the beam.  A beam of frequency $\omega$ can be represented by the
positive frequency part of the electric field operator
$\hat{\mathcal{E}}_{\rm in}^{+} e^{i\omega t}$.  We are interested in
the transverse information of the beam which is fully described by the
slowly varying field envelope operator $\hat{\mathcal{E}}_{\rm
in}^{+}$.  We express this operator in terms of the displaced
TEM$_{n0}$ basis modes, where $n$ denotes the order of the $x$-axis
Hermite-Gauss mode.  Since this paper considers one-dimensional beam displacement, we henceforth denote the beam amplitude function for the transverse modes with only one index.  $\hat{\mathcal{E}}_{\rm
in}^{+}$ can then be written as
\begin{equation}
\hat{\mathcal{E}}_{\rm in}^{+} = i \sqrt{\frac{\hbar \omega}{2
\epsilon_{0}}} \sum_{n=0}^{\infty} \hat{a}_{n} u_{n} (x-d)
\end{equation}
where $u_{n}(x-d)$ are the transverse beam amplitude functions for the
displaced TEM$_{n0}$ modes and $\hat{a}_{n}$ are the corresponding
annihilation operators.  $\hat{a}_{n}$ is normally expressed as
$\hat{a}_{n} = \alpha_{n} + \delta \hat{a}_{n}$, where
$\alpha_{n}=\langle \hat{a}_{n} \rangle$ is the coherent amplitude and
$\delta \hat{a}_{n}$ is a quantum noise operator.  Since the beam is
assumed to be a TEM$_{00}$ mode, only this mode has coherent
excitation and therefore only $\alpha_{0}$ is nonzero.  The field
operator is then given by
\begin{equation} \label{SDin}
\hat{\mathcal{E}}_{\rm in}^{+} = i \sqrt{\frac{\hbar \omega}{2
\epsilon_{0}}} \left( \sqrt{N} u_{0}(x-d) + \sum_{n=0}^{\infty} \delta
\hat{a}_{n} u_{n} (x-d) \right)
\end{equation}
where $\alpha_{0} = \sqrt{N}$.

The difference photo-current, which provides information on the
displacement of the beam relative to the centre of the detector, is
given by
\begin{eqnarray} \label{n1SD}
\hat{n}_{-} & = & \hat{n}_{x>0}-\hat{n}_{x<0} \nonumber\\
& = & \frac{2 \epsilon_{0}}{\hbar \omega} \left[ \int_{-\infty}^{0} dx
(\hat{\mathcal{E}}_{\rm in}^{+})^{\dagger} \hat{\mathcal{E}}_{\rm
in}^{+}-\int_{0}^{\infty} dx (\hat{\mathcal{E}}_{\rm
in}^{+})^{\dagger} \hat{\mathcal{E}}_{\rm in}^{+}\right]
\end{eqnarray}
where $\hat{n}_{x<0}$ and $\hat{n}_{x>0}$ are the photon number operators
for the left and right halves of the detector, respectively.  This
expression can be simplified by changing bases from the TEM$_{n0}$
basis, to a TEM$_{n0}$ basis that has a $\pi$-phase flip at the center of
the detector \cite{fabre}.  This new basis is defined by
\begin{equation}
 v_{n}(x,d) = \left\{ \begin{array}{ll}
 u_{n}(x) & \textrm{for } x > d \\
 - u_{n}(x) & \textrm{for } x < d
 \end{array} \right.
 \label{flipp}
\end{equation}
and we denote annihilation operators for this basis as $\hat b_{n}$.
If the incident TEM$_{00}$ field is bright, so that $N \gg |\langle
\delta \hat{a}_{n}^{2} \rangle|$ for all $n$, the difference
photo-current $\hat{n}_{-}$ can be written compactly as
\begin{equation} 
\hat{n}_{-} = \sqrt{N} \left( \sqrt{N} \zeta_{0} + \delta
\hat{Y}_{0}^{+} \right)
\end{equation}
where $\hat{Y}_{0}^{+}$ is the amplitude quadrature operator
associated with our new flipped basis, given by $\hat{Y}_{0}^{+} =
\langle \hat{Y}_{0}^{+} \rangle + \delta \hat{Y}_{0}^{+} = \hat b_{0}
+ \hat b_{0}^{\dagger}$; and $\zeta_{0}(d) =
\int_{-\infty}^{\infty} v_{0}(x,d) u_{0}(x-d) dx$ is the overlap
coefficient between the first flipped mode and the TEM$_{00}$ mode.
The beam displacement can be inferred from the mean photo-current 
\begin{equation}
\langle \hat{n}_{-} \rangle = N  \zeta_{0},
\end{equation}
where the standard deviation of the photo-current noise is given by $
\Delta \hat{n}_{-} = \sqrt{\langle \hat{n}_{-}^{2} \rangle - \langle
\hat{n}_{-} \rangle^{2}} = \sqrt{N} \Delta \hat{Y}_{0}^{+}.$ For a
coherent field, $\Delta \hat{Y}_{0}^{+} = 1$, making $\Delta
\hat{n}_{-} = \sqrt{N}$.

\begin{figure}[!ht]
\begin{center}
\includegraphics[width=7.5cm]{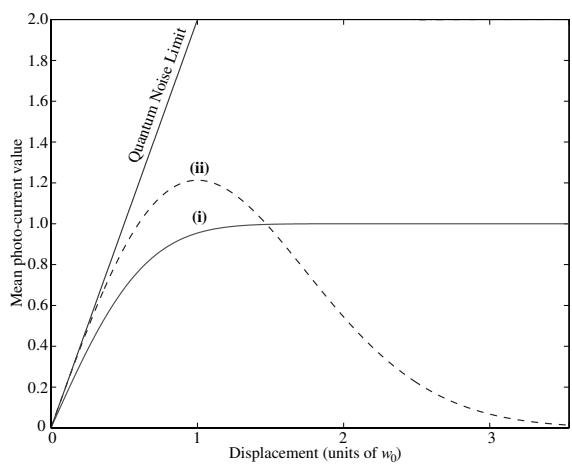}
\caption{Mean value of the normalised difference photo-current,
$\langle \hat{n}_{-} (d) \rangle/N$, as a function of beam
displacement, $d$, for (i) split and (ii) TEM$_{10}$ homodyne detection.}
\label{disp}
\end{center}
\end{figure}
Figure \ref{disp}~(i) shows the normalised difference photo-current
$\langle \hat{n}_{-} \rangle/N$ as a function of beam displacement, $d$, for
split detection.  We see that for small displacements
where $d \ll w_{0}$, the normalised difference photo-current is
linearly proportional to the displacement and can be approximated by
\begin{equation} 
\langle \hat{n}_{-} \rangle_{d \ll w_{0}} = N \zeta_{0, d \ll
w_{0}} \approx \sqrt{\frac{2}{\pi}} \frac{2Nd}{w_{0}}
\label{nsdsmall}
\end{equation}
As $d$ approaches the waist size of the beam, $w_{0}$, the normalised
difference photo-current begins to roll off and asymptotes to a constant 
for larger $d$.  This can be easily understood, since for $d \gg
w_{0}$ the beam is incident almost entirely on one side of the
detector.  In this regime, large beam displacements only cause small
variations in $\langle \hat{n}_{-} \rangle$, making is difficult to
determine the beam displacement precisely.

\begin{figure}[!ht]
\begin{center}
\includegraphics[width=7.5cm]{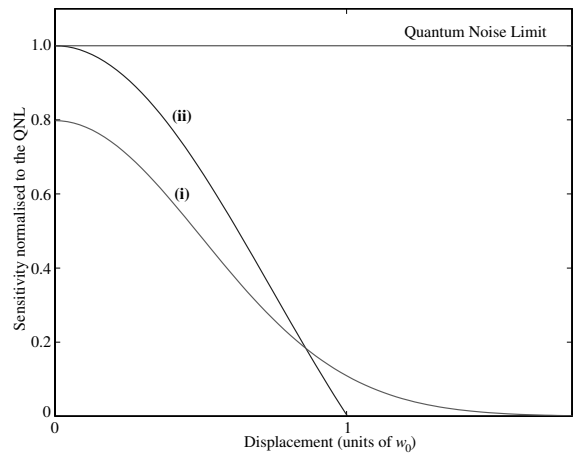}
\caption{Sensitivity response of displacement measurement
for (i) split and (ii) TEM$_{10}$ homodyne detection.}
\label{SNR}
\end{center}
\end{figure}

The noise of our displacement measurement, $\Delta d_{\rm SD}$, is
then related to the noise of the difference photo-current, $\Delta
\hat{n}_{-}$, via
\begin{equation}
\Delta d_{\rm SD} = \frac{\partial d}{\partial \langle \hat{n}_{-}
\rangle} \Delta \hat{n}_{-}
\end{equation}
giving a sensitivity of
\begin{equation}
{\cal S_{\rm SD}} = \frac{1}{\Delta d_{\rm SD}} = \frac{\partial
\langle \hat{n}_{-} \rangle }{ \partial d } \frac{1}{\sqrt{N}} =
\frac{\partial \zeta_{0}}{ \partial d } \sqrt{N}
\end{equation}
for a coherent state. This sensitivity is plotted as a function of displacement in Figure~\ref{SNR}~(i). In the region of small displacement, we have
\begin{equation} 
{\cal S}_{{\rm SD}, d \ll w_{0}} \approx \sqrt{\frac{2}{\pi}} \frac{2 \sqrt{N}}{ w_{0}}
\end{equation}
The efficiency of split detection for small displacement
measurement is therefore given by the ratio
\begin{equation} 
\epsilon_{\rm SD} = \frac{{\cal S}_{{\rm SD}, d \ll w_{0}} }{{\cal
S}_{\rm QNL}}=\sqrt{\frac{2}{\pi}} \sim 80\%
\end{equation}
This $\sqrt{{2}/{\pi}}$ factor arises from the coefficient of the mode
overlap integral, $\zeta_{0}$, between $v_{0}(x,d)$ and $u_{0}
(x-d)$, as shown in Eq.~(\ref{nsdsmall}).  Fig.~\ref{SNR}~(i) shows that
the sensitivity of split detection decreases and asymptotes
to zero for large displacement.  The QNL in the figure confirms
that split detection is not optimal for all displacement.

\subsection{TEM$_{10}$ Homodyne Detection}

\begin{figure}[!ht]
\begin{center}
\includegraphics[width=7.5cm]{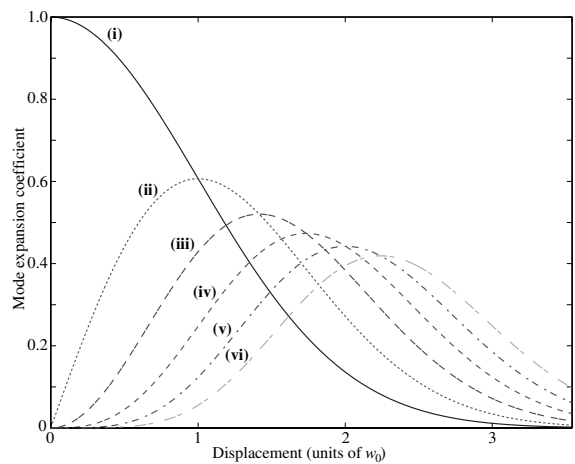}
\caption{Coefficients of the decomposition of the displaced mode in terms of
the TEM$_{n0}$ modes for the (i) TEM$_{00}$, (ii) TEM$_{10}$, (iii)
TEM$_{20}$, (iv) TEM$_{30}$, (v) TEM$_{40}$ and (vi) TEM$_{50}$
mode components.}
\label{cf}
\end{center}
\end{figure}

Before proceeding with our proposal for an optimal small beam
displacement measurement, let us express the displaced TEM$_{00}$ beam
in terms of the centred Hermite-Gauss basis modes, $u_{n}(x)$.
The coefficients of the decomposed basis modes are given by
\begin{equation} \label{coeffs}
\frac{\alpha_{n}}{\sqrt{N}}= \int_{-\infty}^{\infty} u_{0}(x-d)
u_{n}(x)dx = \frac{ d^{n}}{ w_{0}^{n} \sqrt{n!}} \exp \left[ -
\frac{d^{2}}{2 w_{0}^{2}} \right]
\end{equation}
Plots of these coefficients as a function of beam displacement are
shown in Fig.~\ref{cf}.  We notice that for small displacement only
the TEM$_{00}$ and TEM$_{10}$ modes have significant non-zero
coefficients \cite{anderson}.  This means that the TEM$_{10}$ mode
initially contributes most to the displacement signal.  For larger
displacement, other higher order modes become significant as their
coefficients increase.  This suggests that an interferometric
measurement of the displaced beam with a centred TEM$_{10}$ mode
may be optimal in the small displacement regime.

\begin{figure}[!ht]
\begin{center}
\includegraphics[width=6.5cm]{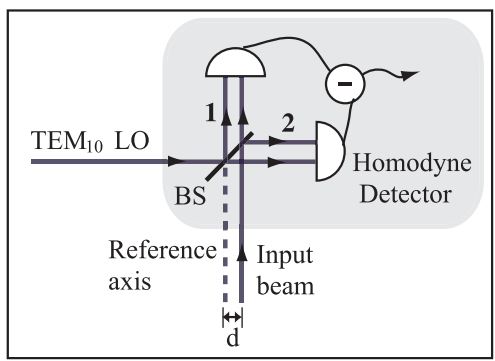}
\caption{TEM$_{10}$ homodyne detection beam displacement measurement. BS: 50/50 beam-splitter, LO: local oscillator.}
\label{scheme}
\end{center}
\end{figure}

Figure~\ref{scheme} shows the TEM$_{10}$ detection scheme considered in this paper, the displaced beam is homodyned with a TEM$_{10}$ mode local oscillator. A reference axis for the displacement of the TEM$_{00}$ beam is
defined by fixing the axis of the local oscillator.  As can be seen from Fig.~\ref{cf}, when the input beam is centred, no power is contained in the TEM$_{10}$ mode. Due to the orthonomality of Hermite-Gauss modes, the TEM$_{10}$ local oscillator beam only detects the TEM$_{10}$ vacuum noise of the input beam. However we see from Fig.~\ref{cf} that displacement of the TEM$_{00}$ input beam couples power into the (centred) TEM$_{10}$ mode. This coupled power interferes with the TEM$_{10}$ local oscillator, causing a change in the photo-current observed by the homodyne detector. Therefore, the homodyne detects a signal proportional to the displacement of the input beam. 

The electric field operator describing the TEM$_{10}$ local oscillator
beam is
\begin{equation} \label{HomoLO}
\hat{\mathcal{E}}_{\rm LO}^{+} = \sqrt{\frac{\hbar \omega}{2
\epsilon_{0}}} \left(\sqrt{N_{\rm LO}} u_{1} (x) + \sum_{n=0}^{\infty}
\delta \hat{a}_{n}^{\rm LO} u_{n}(x) \right)
\end{equation}
where the first bracketed term is the coherent amplitude, the second
bracketed term denotes the quantum fluctuations of the beam, and
$N_{\rm LO}$ is the number of photons in the local oscillator.  Using
this expression and that of the input displaced beam, the photon
number operators corresponding to the two output beams of the
beam-splitter are obtained, given by
\begin{equation}
\hat{n}_{1,2} = \frac{2 \epsilon_{0}}{\hbar \omega} \int_{-\infty}^{\infty}
(\hat{\mathcal{E}}^{+}_{1,2})^{\dagger} \hat{\mathcal{E}}^{+}_{1,2} dx
\end{equation}
where subscripts $\{1,2\}$ denote the two output beams.  From these, the
difference photo-current between the two detectors used for homodyning is
\begin{eqnarray} \label{nhomo}
\hat{n}_{-} & = & \sqrt{N_{\rm LO}} (2 \alpha_{1} + \delta
\hat{X}_{1}^{+}) \nonumber\\
& = & \sqrt{N_{\rm LO}} \left( \frac{2 \sqrt{N}}{ w_{0}}d + \delta
\hat{X}_{1}^{+} \right)
\end{eqnarray}
where $\delta \hat{X}_{1}^{+} = \delta \hat{a}_{1} + \delta
\hat{a}_{1}^{\dagger}$ is the amplitude quadrature noise operator of the TEM$_{10}$ component of the input beam, and we have assumed that $N_{\rm LO} \gg N$.

The mean photo-current as a function of beam displacement is shown in
Figure~\ref{disp}~(ii).  For small displacement, the mean photo-current is
linearly proportional to the beam displacement.  The factor of
proportionality, $2 \sqrt{N}/ w_{0}$, in this case is the same as that
of the array detection scheme.  Hence, this shows that the
measurement is optimal.  In the large displacement regime, the
photo-current decreases to zero.  This is because the displacement
signal couples to higher order modes.

The sensitivity response for the TEM$_{10}$ detection, obtained in the same manner as that for split detection, is shown in Figure~\ref{SNR}~(ii).  In the small displacement regime, we obtain
\begin{equation} 
{\cal S}_{H,d \ll w_{0}} = \frac{2 \sqrt{N}}{ w_{0}}
\end{equation}
The efficiency of the TEM$_{10}$ detection is then
\begin{equation} 
\epsilon_{H} = \frac{{\cal S}_{H,d \ll w_{0}}}{{\cal S}_{\rm QNL}} = 100\%.
\end{equation}
For larger displacement, the sensitivity decreases as the displacement
increases.  This is due to the power of the displaced beam being
coupled to higher-ordered TEM$_{n0}$ modes (for $n > 1$) and thus less
power is contained in the TEM$_{10}$ mode.

%%%%%%%% Squeezing %%%%%%%%
\section{Displacement measurement beyond the quantum noise limit}

\subsection{Split Detection with Squeezed Light}

It has been shown that the noise measured by split detection is the
flipped mode defined in Eq.~(\ref{flipp}) \cite{fabre}. In order to improve the sensitivity of split detection, squeezing has to be introduced on the flipped
mode.  This method was demonstrated in the initial work of Treps {\it
et al.} and the resulting spatial correlation between the two halves of the beam was termed {\it spatial squeezing} \cite{treps1d}.  By applying a
displacement modulation to the spatially squeezed beam, they
demonstrated that sensitivities beyond the QNL could be achieved for beam displacement measurements. Treps {\it et al.} recently extended their one-dimensional spatial squeezing work to the two orthogonal transverse spatial axes \cite{treps2d, treps2da}.  In this scenario, the photon correlation was measured between the set of top and bottom halves as well as left and right halves of a quadrant detector.

Restricting our analysis to one dimension, Fig.~\ref{sqz}~(a) shows the
combination of an input TEM$_{00}$ beam with a vacuum squeezed
symmetric flipped mode $v(x,0)$ as defined in Eq.~(\ref{flipp}).
Lossless combination of orthogonal spatial modes can be achieved in a
number of ways.  One example is the use of optical cavities for mixing
resonant and non-resonant modes as illustrated by the Figure.  The
cavity reflects off the vacuum squeezed symmetric flipped mode whilst
transmitting the TEM$_{00}$ beam.  The total beam is then displaced
and measured using a split-detector.  The beam incident on the split
detector is described by
\begin{eqnarray} \label{sqzSD}
\hat{\mathcal{E}}_{\rm in}^{+} & = & i\sqrt{\frac{\hbar \omega}{2
\epsilon_{0}}} \Big( \sqrt{N} u_{0}(x-d) + \delta \hat{c}_{0}^{\rm
sqz} v_{0}(x-d,0) \nonumber\\
& & + \sum_{n=1}^{\infty} \delta \hat{c}_{n} v_{n}(x-d,0) \Big)
\end{eqnarray}
where the first term arises from the TEM$_{00}$ mode while the second
term describes the vacuum squeezed flipped mode and the last term in
Eq.~(\ref{sqzSD}) represents all other higher order vacuum noise
terms.  The photo-current difference operator in the limit of small
displacement is
\begin{equation}
\hat{n}_{-} = \sqrt{N} \left( \sqrt{\frac{2}{\pi}} \frac{2\sqrt{N}}{
w_{0}} d+ \delta \hat{Z}_{0,{\rm sqz}}^{+} \xi_{0} +
\sum_{n=1}^{\infty} \delta \hat{Z}_{n}^{+} \xi_{n} \right)
\end{equation}
where $\delta \hat{Z}_{n}^{+} = \delta \hat{c}_{n} + \delta
\hat{c}_{n}^{\dagger}$ are the amplitude quadrature operators.  The
first bracketed term arises from the mode overlap between the $v_{0}
(x,0)$ and $u_{0}(x-d)$ modes, which has a value of $\sqrt{2/\pi}$.
The second term originates from the mode overlap between the
$v_{0}(x,d)$ and $v_{0}(x-d,0)$ modes.  The last term is a result of
the overlap between the $v_{0} (x,d)$ and $v_{n}(x-d,0)$ modes.  The
overlap coefficients are given by $\xi_{n}
=\int_{-\infty}^{\infty} v_{0} (x,d) v_{n}(x-d,0) dx$.

Figure~\ref{slopesqz}~(i) plots the sensitivity, $\cal{S}_{\rm SD}$, as a
function of squeezing.  Notice that the QNL can only be surpassed with more than 1.9~dB of squeezing. This is a direct consequence of the intrinsic inefficiency of displacement measurement using split detection. 

\subsection{TEM$_{10}$ Homodyne Detection with Squeezed Light}

\begin{figure}[!ht]
\begin{center}
\includegraphics[width=7.5cm]{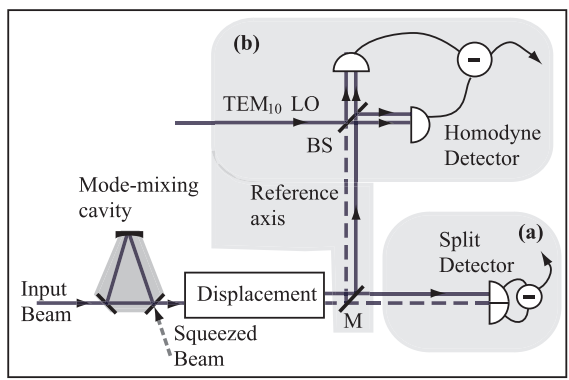}
\caption{Measurement options: (a) Split detection with a vacuum squeezed symmetric flipped mode or (b) TEM$_{10}$ homodyne detection with a vacuum squeezed TEM$_{10}$ mode.  Both schemes combine the squeezed beam and the TEM$_{00}$ input beam losslessly using an optical cavity. M: mirror, BS: 50/50 beam-splitter, LO: local oscillator.}
\label{sqz}
\end{center}
\end{figure}
We have shown that by squeezing the flipped mode and detecting the
beam displacement using a split detector, one is able to improve on
the displacement sensitivity to beyond the QNL. Correspondingly, we now
consider the effect of squeezing the TEM$_{10}$ mode in our homodyne
detection.  Figure~\ref{sqz} shows the combination of the
TEM$_{00}$ input beam with a vacuum squeezed TEM$_{10}$ beam prior to
displacement.  The displaced beam is then analysed using homodyne
detection with a TEM$_{10}$ local oscillator beam as discussed
previously (see Fig.~\ref{sqz}~(b)).  The detected beam is given by
\begin{eqnarray}
\hat{\mathcal{E}}_{\rm in}^{+} & = & i\sqrt{\frac{\hbar \omega}{2
\epsilon_{0}}} \Big( \sqrt{N} u_{0} (x-d) + \delta \hat{a}_{1}^{\rm
sqz} u_{1}(x-d) \nonumber\\
& & + \sum_{n \neq 1}^{\infty} \delta \hat{a}_{n} u_{n}(x-d) \Big) 
\end{eqnarray}
where the first term arises from the TEM$_{00}$ mode, the second
term from the squeezed vacuum TEM$_{10}$ mode, and
the last term from higher ordered vacuum noise.  The
difference photo-current between the two detectors used for homodyning is 
given by
\begin{equation} \label{nsdsqz}
\hat{n}_{-} = \sqrt{N_{LO}} \Big( \frac{ 2 \sqrt{N}}{ w_{0}}d +\delta
\hat{X}_{1,{\rm sqz}}^{+} \chi_{1} + \sum_{n \neq 1}^{\infty}
\delta \hat{X}_{n}^{+} \chi_{n} \Big)
\end{equation}
where $\chi_{n} = \int_{-\infty}^{\infty} u_{1} (x) u_{n} (x-d) dx$.  For small displacement, the overlap between the vacuum squeezed mode and the local oscillator beam is $\sim 100\% $ whilst the last term is
negligible.  

\begin{figure}[!bt]
\begin{center}
\includegraphics[width=7.5cm]{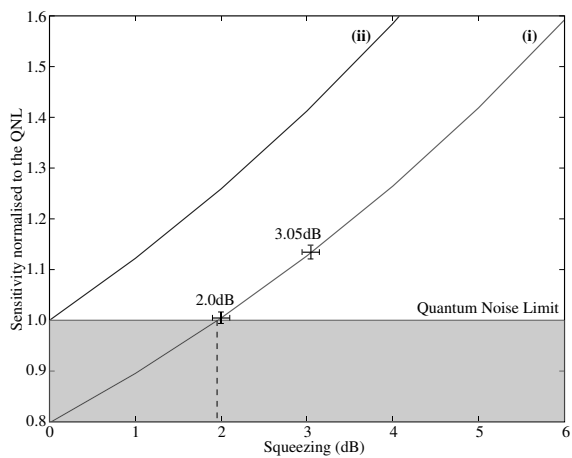}
\caption{Plots of sensitivity, $\cal{S}$, for small displacement, as a
function of squeezing for (i) split and (ii) TEM$_{10}$ homodyne detection.  Shading indicates the region where displacement measurement is below the QNL. The data points were obtained from Ref.~\cite{treps2d}.}
\label{slopesqz}
\end{center}
\end{figure}

The sensitivity, $\cal{S}_{H}$, as a function of squeezing on the
TEM$_{10}$ mode is shown in Fig.~\ref{slopesqz}~(ii). Since the scheme is optimum for small displacement, any amount of squeezing will
lead to a sensitivity beyond the QNL. Furthermore, the TEM$_{10}$ detection surpasses the performance of split detection for all values
of squeezing.

\subsection{Discussion}

In the paper on quantum displacement measurement by Treps {\it et al.}
\cite{treps2d}, displacement measurements of the two transverse axes
were performed using split detection with two co-propagating
squeezed beams.  The squeezing values were 2.0~dB and 3.05~dB for the
vertical and horizontal displacement measurement, respectively.
Relating this result back to our analysis, we find that the
displacement measurements performed were indeed beyond the QNL. These
corresponded to sensitivities of 100.5\% and 113.0\% above the QNL as shown in Fig.~\ref{slopesqz}.  However, using the same squeezed beams but adopting the TEM$_{10}$ homodyne detection, sensitivities of 126\%
and 141.5\% above the QNL would be achievable. 

The TEM$_{10}$ homodyne detection can be extended to perform beam
displacement measurements in both transverse dimensions. The TEM$_{10}$ mode component is responsible for beam displacement in one transverse axis. Thus, by symmetry, the TEM$_{01}$ mode component is responsible for beam displacement in the orthogonal transverse axis. Correspondingly, beam displacement measurement in the horizontal or vertical transverse axis can be achieved by adapting the mode-shape of the local oscillator beam to either TEM$_{10}$ or TEM$_{01}$. To perform optimum larger beam displacement measurements, the local oscillator of the TEM$_{10}$ homodyne detection can be modified by including higher order components of the TEM$_{00}$ displaced beam.  Similarly, the homodyning scheme can be extended to measuring displacements of arbitrary mode-shapes assuming {\it a priori} knowledge of the beam shape.

%%%%%%%% Conclusion %%%%%%%%
\section{Conclusion}

By defining the beam position as the mean photon position of a light
beam, optical beam displacement can be measured with reference to a
fixed axis.  Using an idealised array detection scheme, we derived the
QNL associated with optical displacement measurements.  A displaced
TEM$_{00}$ beam can be decomposed into an infinite series of
Hermite-Gauss modes but in the limit of small displacement, only the
TEM$_{00}$ and TEM$_{10}$ components are non-negligible.  Since the
split detector effectively measures the noise of an optical flipped
mode \cite{fabre}, it is only $\sim 80\%$ efficient when used to
measure the displacement of a TEM$_{00}$ beam.  We have proposed an
optimum displacement measurement scheme based on homodyne detection. By using a TEM$_{10}$ local oscillator, small displacement signals can be extracted with 100\% efficiency. We showed that in this small displacement regime the TEM$_{10}$ homodyne detection performs at the QNL, and is significantly more efficient than split detection. 

We have also shown that by mixing the input beam with a squeezed beam
in the appropriate mode, we can significantly improve the sensitivity
of the TEM$_{10}$ homodyne detection.  We compared the sensitivities of
both split and TEM$_{10}$ homodyne detection for equal values
of squeezing and found that for small displacements the TEM$_{10}$ detection outperforms split detection for all values of squeezing. For split detection, more than 1.9~dB of squeezing is required to achieve a sensitivity beyond the QNL. Whilst for the TEM$_{10}$ detection any amount of squeezing will suffice.

%%%%%%%% Acknowledgements %%%%%%%%
\section{Acknowledgment}
We thank Nicolai Grosse and Nicolas Treps for fruitful discussions.
This work is funded by the Australian Research Council Centre of Excellence Program.

%%%%%%% References %%%%%%%%

\end{document}